\documentstyle[11pt,newpasp,twoside,epsf]{article}
\def\edcomment#1{\iffalse\marginpar{\raggedright\sl#1\/}\else\relax\fi}
\reversemarginpar

\begin{document}
\title{A direct imaging search for wide (sub)stellar companions around rad-vel planet host stars}
 \author{M.Mugrauer, R.Neuh\"auser, T.Mazeh, M.Fernandez, E.Guenther and C.Broeg}
 \affil{AIU, Schillerg\"asschen 2-3, 07745 Jena}

\begin{abstract}
We present an overview of our ongoing systematic search for wide (sub)stellar companions around the
stars known to host rad-vel planets. By using a relatively large field of view and going very deep
our survey can find all directly detectable stellar and brown dwarf companions (m$>$40\,$M_{Jup}$)
within a 1000\,AU orbit.
\end{abstract}

As of may 2003 more than one hundreds extra-solar were found. Some of those planets are found in
binary stellar systems. Those cases are intriguing, and might even exhibit some statistical
different features than the planets around single stars (Zucker \& Mazeh 2002). It is therefore
that we started a systematic deep imaging of the stars known to have planets to look for faint
companions in wide orbits. Several groups have already searched for very close stellar companions
using adaptive optics, but these searched could not find wide companions due to their too small
field of view. By using a relatively large field of view and going very deep our survey will find
all stellar and even brown dwarf companions (m$>$40\,$M_{Jup}$) within a 1000\,AU orbit.\newline

We use the ESO 3.58-m NTT, the 3.8-m UKIRT and the 2.2-m Calar Alto telescopes to image all stars
known to host planets.  These stars are relatively close and therefore their proper motion is
large. Consequently, their 1$^{st}$ epoch observations can be compared with images from 2MASS to
reveal objects that co-move with the rad-vel planet host stars. Objects brighter than
H$\sim$15\,mag (2MASS limit) can be studied using this method (see Figure\,1). To reach the
detection limit of the NTT, which is at H$>$19\,mag, we need 2$^{nd}$ epoch NTT observations. By
this approach we can find faint and therefore low-mass companions even with planetary masses. Due
to a pixelscale which is more than 10 times better than 2MASS, an epoch difference of almost one
year is sufficient to find co-moving companions around nearly all rad-vel planet host stars (see
Figure\,2).

\begin{figure}[hbt]

\plotone{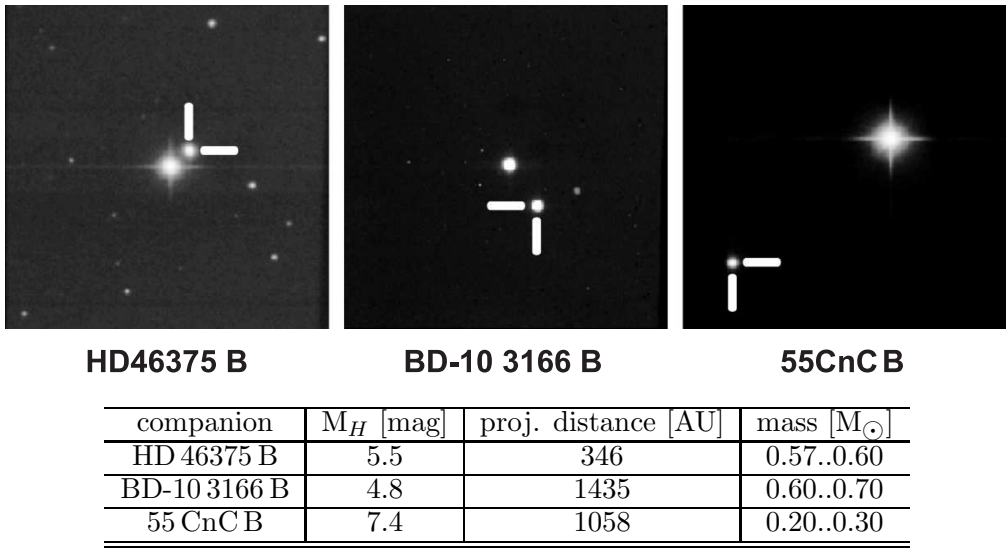} \newline\newline
\begin{tabular}{c|c|c|c}
\tableline
companion          & M$_{H}$ [mag] & proj. distance [AU]   & mass [M$_{\sun}$] \\
\hline
HD\,46375\,B    & 5.5           & 346                   & 0.57..0.60 \\
\hline
BD-10\,3166\,B  & 4.8           & 1435                  & 0.60..0.70 \\
\hline
55\,CnC\,B      & 7.4           & 1058                  & 0.20..0.30 \\
\tableline\tableline
\end{tabular}

\caption{H band images obtained with MAGIC at the CA~2.2m. The rad-vel planet host stars HD\,46375,
BD-10~3166 and 55\,CnC are shown with their co-moving stellar companions (see white markers) which
were detected by comparing the MAGIC images with 2MASS images. The absolute H magnitude, the
projected distance to the primary and the estimated mass of the stellar companions are shown in the
table below.}
\end{figure}

\begin{figure}[hbt]
\plotone{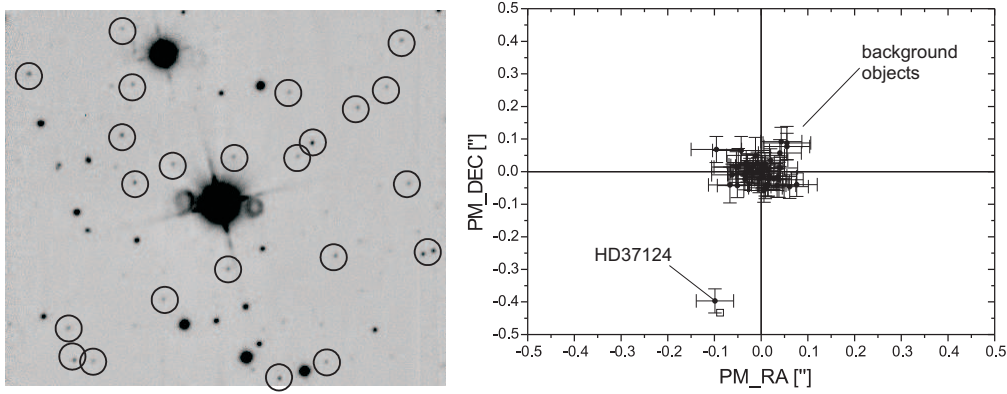} \caption{Left: NTT H band image of the rad-vel planet host star
HD\,37124. Objects in black circles are not detected in 2MASS. By comparing two NTT images taken
with an epoch difference of one year we could measure the spatial motion of all detected faint
companion-candidates. Right: Derived proper motion of objects detected on the left image. The white
square is the expected motion of HD\,37124 for the given epoch difference, based on Hipparcos data.
All faint objects are clearly not co-moving, hence they are all background stars.}
\end{figure}

\end{document}